\documentclass[10pt]{iopart}

\usepackage[T1]{fontenc}
\usepackage[utf8]{inputenc}
\usepackage{graphicx}

\usepackage{iopams}  
\expandafter\let\csname equation*\endcsname\relax
\expandafter\let\csname endequation*\endcsname\relax
\usepackage{amsmath}

\usepackage{varioref}
\usepackage{hyperref}
\hypersetup{
    hidelinks
}
\usepackage{cleveref}
\usepackage[english]{babel}
\usepackage{float}
\usepackage{xcolor}
\usepackage{cite}
\usepackage{orcidlink}

\begin{document}

\title[]{A new model for runaway electron transport based on chaotic Hamiltonian systems}

\author{Dániel Jánosi$^{1,2,3}$\orcidlink{0000-0002-6138-3610}, Anikó Horváth$^1$\orcidlink{0009-0009-4502-2397}, Hannes Bergström$^4$\orcidlink{0009-0004-6274-1661}, Matthias Hölzl$^{4,5}$\orcidlink{0000-0001-7921-9176}, Gergely Papp$^4$\orcidlink{0000-0003-0694-5446}, Gábor Veres$^{3,6}$, Gergo I. Pokol$^{3,6}$ and György Károlyi$^3$\orcidlink{0000-0002-1021-9554}}
\address{$^1$Department of Theoretical Physics, Eötvös Loránd University, Pázmány Péter Sétány 1/A, H-1117 Budapest, Hungary}
\address{$^2$HUN-REN Institute of Earth Physics and Space Science, Csatkai Endre utca 6-8, H-9400 Sopron, Hungary}
\address{$^3$Institute of Nuclear Techniques, Faculty of Natural Sciences, Budapest University of Technology and Economics, Műegyetem rakpart 3, H-1111 Budapest, Hungary}
\address{$^4$Max Planck Institute for Plasma Physics, Boltzmannstrasse 2, D-85748 Garching, Germany}
\address{$^5$Department of Physics and Astronomy, Chalmers University of Technology, Göteborg, SE-41296, Sweden}
\address{$^6$HUN-REN Centre for Energy Research, Department of Fusion Technology, Konkoly-Th. M. 29-33, H-1121 Budapest, Hungary}

\vspace{10pt}

\begin{abstract}

\noindent
The transport of runaway electrons (RE) in ergodic magnetic geometries is an area of active study. Computing the transport from the direct simulation of particle trajectories is computationally expensive. Instead, diffusion models, such as the one by Rechester and Rosenbluth, are often employed to incorporate transport effects into reduced simulations. However, the comparison of diffusion-based to direct simulations reveals that the transport is typically not purely diffusive. In this paper, we introduce a simple transport model, based on chaos theory, which goes beyond the Rechester-Rosenbluth approximation. Besides chaotic diffusion, our model takes into account the effect of so-called sticky regions, a trapping layer around magnetic islands, where particle escape slows down to a power-law decay rather than an exponential decay. We demonstrate the applicability of the model both in the Ullmann-Caldas map with parameters corresponding to the TBR-1 tokamak, and in a JOREK simulation of a JET disruption scenario, with remarkably good fits achieved in both cases.

\end{abstract}

\vspace{2pc}

\submitto{\PPCF}
%
%
\ioptwocol

\section{Introduction}

Disruptions are a critical outstanding issue on the path to controlled nuclear fusion power plants of the tokamak type~\cite{Hender_2007_ITER_Physics, Lehnen_2015_ITER, Hollmann_2015_Status}. Beyond thermal loads and electromagnetic forces, the generation of a relativistic runaway electron (RE) beam is of great concern~\cite{breizman2019physics}. Runaways can cause serious localised damage~\cite{Matthews_2016} and have the potential for deep melting of structures behind plasma-facing components. For this reason, significant effort is expended on the research of runaway electron dynamics, avoidance, and mitigation~\cite{PazSoldan_2021, Reux_2022, Sheikh_2024}. 

The runaway avalanche process -- leading to exponential multiplication of the number of relativistic electrons -- is the main cause for the potential generation of high-current runaway-electron beams~\cite{Rosenbluth_1997, Hesslow_2019a}. The avalanche process, however, relies on a primary seed population, which can emerge via the Dreicer~\cite{Dreicer_1959,Dreicer_1960,Hesslow_2019b}, hot-tail~\cite{smith08hot}, or nuclear processes~\cite{MartinSolis_2017_RE}. For most present-day tokamaks, as well as in modelling for future tokamaks such as ITER, the runaway plateau current is dominated by avalanche-generated runaways~\cite{Pautasso_2016, Pautasso_2020, Insulander_2020_Kinetic, Insulander_2021_Modelling, linder2020self, linder2021electron, Vallhagen_2024}.

The seed population, however, may not always survive the MHD mixing and enhanced transport during the quench phase of the disruption. Present day experiments aimed at studying REs do not always observe a stable RE beam formation, even if the scenario is repeated precisely~\cite{Pautasso_2016, Pautasso_2020, decker2022full, heinrich2024characteristics}. One of the possible causes for this is the loss of the seed population, upon which the exponentiation of RE avalanche depends~\cite{Boozer_2016,Linder_2020,Linder_2021,Insulander_2021,Hoppe_2021_AUG,Svensson_2021,gobbin2017runaway,gobbin2021role,gobbin24runaway}.

The quench phase and the transport that arises during it are remarkably difficult to study both experimentally and by modelling. One of the highest fidelity approaches for numerical study is using tracer test particles in 3D non-linear magnetohydrodynamic (MHD) simulations~\cite{Sommariva_2018,Sarkimaki_2022}. However, the drawback of this is the required amount of computational power, which limits the number of scans that can be executed and limits the phase space that can be covered by parameter scans. In future tokamaks -- like ITER -- the nuclear sources are expected to provide a continuous seed of primary REs, combined with high avalanche multiplication potential~\cite{MartinSolis_2017_RE,Vallhagen_2020,Vallhagen_2024}. Nevertheless, the understanding of RE transport, and in particular RE seed transport is a key requirement for better interpretation of present-day RE experiments, which is crucially important for RE model validation.

A typical approximation of RE transport in volumes with ergodic (i.e. chaotic) or stochastic fields is via the Rechester-Rosenbluth (RR) diffusion~\cite{rechester78electron}. This fundamental result provides modellers with the opportunity to incorporate the effect of magnetic field perturbations with a theoretically sound model that has a single control parameter, the $\delta B / B$ relative perturbation strength. However, the model assumes an enclosed ergodic volume, and its applicability may be limited in cases where the ergodic volume is open to the wall on one side~\cite{papp2011runaway, papp2011ITER, papp2012effect, papp2015energetic}. Simulation studies have shown that under such circumstances, the RR diffusion is insufficient to describe the RE transport observed in test particle simulations~\cite{papp2015energetic, Sarkimaki_2020}. RR diffusion also assumes a fully ergodic volume, but the underlying magnetic topology can be more complex.

In this paper, we exploit the correspondence between phase space topology of Hamiltonian systems and magnetic field lines. This correspondence sheds light on the fact that beyond intact surfaces and fully ergodic volumes, there exist regions -- referred to as ``sticky regions'' -- where the transport is significantly slower than in the rest of the volume. We propose a novel semi-analytical model to describe RE transport in such complex perturbed fields, going beyond the RR description. The model is derived from chaotic transport, and accounts for the effect of the sticky regions~\cite{percival1980,tel2011,karney1983,ott1993,Janosi_2019,janosi2024a,JT24}. We show that this functional form can successfully describe the temporal evolution of particle confinement probabilities in both a simple approximative mapping, as well as in a complex simulation of test particles in a 3D non-linear MHD simulation with JOREK~\cite{Nardon_2021,Hoelzl_2021,Nardon_2023,Hoelzl2024recent,Bandaru_2024,Vannini_2025,Hoelzl_2026}. 

The manuscript is organised as follows. We introduce the transport model in section~\ref{sec:model}, followed by a description of the Ullmann-Caldas map in section~\ref{sec:UCmap}. In section~\ref{sec:JET} we demonstrate the model on a JET RE scenario, followed by a discussion and conclusions.

\section{The transport model \label{sec:model}}

When a Hamiltonian, integrable system is perturbed with some non-linear effect, some invariant curves break up, leaving narrow bands of irregular behaviour in the phase space. These bands in turn act as separatrices, between which islands can form, possessing irrational safety factors -- or winding numbers, in the language of dynamical systems theory. If the perturbation is increased, then the narrow irregular bands between islands grow, giving rise to macroscopic chaotic regions, called chaotic seas. This coexistence of islands and chaotic seas is called a mixed phase space in chaos theory, a typical feature of chaotic Hamiltonian systems~\cite{ott1993,tel2006} as well as magnetic field lines in tokamaks, see e.g.~\cite{Portela2008}. For a given value of the perturbation, the dynamics within the islands is quasi-periodic, meaning that trajectories visit the vicinity of their initial conditions almost periodically (exact periodicity is observed for rational safety factors), always staying within a well-defined closed curve. By contrast, inside the chaotic sea, the basic properties of chaos can be observed: the trajectories are irregular, i.e. completely aperiodic, and are sensitive to initial conditions. The latter property means that two trajectories initiated arbitrarily close to each other will diverge exponentially, and eventually end up far from each other. The average rate of this divergence is called the Lyapunov exponent, and it is always positive in chaos~\cite{ott1993,tel2006}.

In mixed phase spaces there exists another type of region, found at the edge of the islands, and consists of the remnants of the outermost islands, broken up due to an increase in perturbation~\cite{percival1980}. A chaotic trajectory is able to enter such a region, and when it does, it gets trapped inside the complicated structure of island remains, and can spend a long time there before escaping back to the chaotic sea. Because of this feature, these areas are often called {\em sticky regions}, as here trajectories ``stick'' to the outside of islands. The sticky regions are open towards the chaotic sea, but do not possess chaotic dynamics themselves, in fact, they ``slow down'' chaotic trajectories by trapping them near the island.

There are several studies linking the behaviour of magnetic field lines of tokamaks to two-dimensional Hamiltonian maps~\cite{Jakubowski2006,portela2007,Portela2008,souza2024}. One of the reasons this connection can be made is that considering the position of field lines on a poloidal surface constitutes a Poincaré section, a standard method of investigating Hamiltonian dynamics. This way, one is able to define simple mapping rules and study the complex behaviour of field lines on a two-dimensional plane. A non-linear mapping usually results in a mixed phase space, with the chaotic sea typically showing up at the outer regions of the plasma domain. A number of such non-linear maps exist~\cite{ULLMANN2000,martin1984,abdullaev96application,abdullaev98twist,balescu98tokamap}. The applicability of these maps is of course limited to the behaviour of field lines, or to particles obeying the field line tracing approximation. Since in this area there is essentially a one-to-one correspondence between the concepts of plasma physics and dynamical systems theory (e.g. safety factor -- winding number), in the following we will primarily use the former in order to avoid unnecessary jargon.

Chaotic transport is a topic which has been extensively studied within chaos theory~\cite{ott1993,altmann2013,tel2011}. Of particular interest to us is the case of mixed phase space, i.e., when both chaotic seas and magnetic islands are present. Then, two effects can influence the survival probability, i.e., the probability $P$ that particles (in the field line tracing approximation) do not escape by a given time $t$. First, within the chaotic sea, this probability decays exponentially, $P(t) \sim e^{-\kappa t}$, where $\kappa$ is the escape rate~\cite{tel2011}. Second, the presence of islands, themselves barriers to transport, results in the emergence of sticky regions, where the decay is slower. Earlier investigations pointed out that the time spent in sticky regions~\cite{channon1980}, the survival probability~\cite{karney1983}, as well as other quantities~\cite{meiss1983,chirikov1984} behave as a power-law, rather than having an exponential decay.

The transport model proposed by us, based on the theoretical model of Jánosi and Károlyi~\cite{janosi2024}, has the core concept that since particles can be found in both the chaotic and sticky regions at all times, both of their contributions to transport have to be taken into account at all times. That is, the survival probability has to possess two terms, describing the exponential and power-law decays. 

We assume one-dimensional -- in practice radial -- or layer-to-layer transport, without any particle source aside from the initial distribution. Out of the total number $N_\text{tot}(r,t)$ of particles at a given radius, $N_\text{island}(r)$ fall into an island and thus do not contribute to the transport. The number of those that do contribute, the ones falling into the chaotic and sticky regions, respectively, are denoted as 
\begin{align}
    N(r,t) =& N_\text{tot}(r,t) - N_\text{island}(r) \nonumber \\ 
    =& N_\text{chaotic}(r,t) + N_\text{sticky}(r,t).
\end{align}
The time-dependence of the latters, based on the considerations above, can be written as
\begin{equation}
    N_\text{chaotic}(r,t) \sim e^{-\kappa t}, \;\;\;\;\; N_\text{sticky}(r,t) \sim \frac{1}{1+\big(\frac{t}{\tau}\big)^{\alpha}},
\end{equation}
where $\tau$ is the time-scale associated with the power-law behaviour with exponent $\alpha$, and the 1 in the denominator is written so that the formulas are valid for $t = 0$. Note that we made the assumption that none of the parameters depend on either the radius or time, which are reasonable assumptions for a typical mixed phase space, low-dimensional chaotic system. We say that the ratio of chaotic and sticky particles, respectively, at $r$, is 
\begin{equation}
    \frac{N_\text{chaotic}(r,t)}{N(r,t)} = A(r), \;\;\;\;\; \frac{N_\text{sticky}(r,t)}{N(r,t)} = 1 - A(r).
\end{equation}
Here again we assumed time-independence; for further details on these assumptions, see the original derivation~\cite{janosi2024}. We thus write
\begin{equation}
    N(r,t) = N(r,0)A(r)e^{-\kappa t} + N(r,0)\frac{1 - A(r)}{1+\big(\frac{t}{\tau}\big)^{\alpha}},
\end{equation}
where the appearance of the initial condition in both terms implies the simultaneous presence of both terms at all times. 

Then, we integrate over $r$, obtaining the {\em statistical transport function}
\begin{equation}
    P(t) = Ae^{-\kappa t} + \frac{1 - A}{1+\big(\frac{t}{\tau}\big)^{\alpha}},
    \label{eq:transport_function}
\end{equation}
describing a coarse-grained behaviour of particles, on the scale of the vessel. Here the survival probability is defined, by dividing with the global initial condition $N(0)$, as 
\begin{equation}
    P(t) = \frac{N(t)}{N(0)} = \frac{N_\text{tot}(t) - N_\text{island}}{N_\text{tot}(0) - N_\text{island}},
    \label{eq:P_def}
\end{equation}
with $N(t)$ being the integrated particle number, and the ratios are obtained through the weighted average
\begin{equation}
    A = \frac{1}{N(0)}\int{A(r)N(r,0)\text{d}r}.
    \label{eq:A_avg}
\end{equation}

Although both terms are present at all times, their relative importance is not constant. Typically, the exponential decay dominates initially, while the power-law shows up as a prolonged tail. To give a rough estimate for when the crossover between the two regimes happens, one can equate the terms on the right hand side of \eqref{eq:transport_function} to obtain 
\begin{equation}
    -\kappa t_x + \ln{\frac{A}{1 - A}} = -\ln{\left[1 + \left(\frac{t_x}{\tau}\right)^{\alpha}\right]},
    \label{eq:crossover}
\end{equation}
where $t_x$ is the crossover time. This is a transcendent equation and can only be solved numerically after obtaining the values of the parameters.

The validity of this transport model was successfully demonstrated on the standard map~\cite{janosi2024}, one of the simplest chaotic maps representing field lines~\cite{Escande_2016}. It is important to note that there have already been models in the literature which take into account both the exponential and the power-law effects~\cite{altmann2013}. The unique feature of this model is the presence of both terms at all times, as some models prescribe time intervals for the validity of the terms~\cite{tel2011}, and on the other hand the presence of the distinct time-scale $\tau$ for the power-law decay, which is either not found in other models in the literature, or the time-scales of the two regimes are assumed to be the same~\cite{altmann2008,altmann2013}.

\section{Ullmann-Caldas map \label{sec:UCmap}}

\subsection{The model}

One of the simple maps describing the behaviour of magnetic field lines as defined on a Poincaré section was introduced by Ullmann and Caldas \cite{ULLMANN2000}. The model is a composition of two maps: one describing the evolution of field lines in a large aspect ratio tokamak - using the cylindrical coordinate approximation -  at equilibrium, and the other the effect of an ergodic magnetic limiter as a perturbation of the equilibrium state. 
The equilibrium mapping is given as 
\begin{align}
    r_{n+1}^{\rm eq} &= \frac{r_n}{1 - a_1 \sin \theta_n}, \nonumber \\
    \theta_{n+1}^{\rm eq} &= \theta_n + \frac{2\pi }{q(r_{n+1}^{\rm eq})} + a_1 \cos \theta_n, 
    \label{eq:UC_eq}
\end{align}
where $r_n$ and $\theta_n$ are the radial and poloidal coordinates on the Poincaré section after $n$ steps. Coefficient $a_1 = -0.04$ comes from the lowest-order term of an infinite series representing the toroidal correction to the cylindrical approximation, and is sufficient in accounting for such effects in this map \cite{ULLMANN2000}. The term $q(r)$ is the safety factor profile, for which the following form is used, typically observed in experiments \cite{portela2007}:
\begin{equation}
    q(r) = q_a \frac{r^2}{a^2} \cdot \frac{1}{1- \Theta(a-r)   \left( 1 +\beta' \frac{r^2}{a^2} \right)  \left( 1 - \frac{r^2}{a^2} \right)^{2} },
\end{equation}
where $q_a =3.9$ is the safety factor at the edge of the plasma, $a$ is the plasma minor radius, $ \beta'=1$ is an experimental value (for details see Portela et al.~\cite{portela2007}), and $\Theta$ is the Heaviside step function. Plots of this safety factor profile are available in the references~\cite{portela2007,ULLMANN2000,janosi2024a}. 

The effect of the ergodic magnetic limiter is represented by another mapping acting on \eqref{eq:UC_eq} as
\begin{align}
        \theta_{n+1} &= \theta_{n+1}^{\rm eq} - C \left( \frac{r_{n+1}^{\rm eq}} {b} \right)^{m-2} \cos (m\theta_{n+1}^{\rm eq}), \nonumber \\
         r_{n+1}^{\rm eq} &= r_{n+1} + \frac{mCb}{m-1} \left( \frac{r_{n+1}} {b} \right)^{m-1} \sin (m\theta_{n+1}^{\rm eq}),
         \label{eq:UC_lim}
\end{align}
where $m = 3$ is the poloidal mode number and $b$ is the vessel minor radius. Coefficient $C$ is defined as 
\begin{equation}
    C = \frac{2mla^2}{R_0q_ab^2}\varepsilon,
\end{equation}
where $R_0$ is the major axis radius, $l$ is the limiter thickness, and $\varepsilon = I_l/I_p$ is the ratio of the limiter- and plasma currents, the control parameter of the model. It is apparent that Eq.\eqref{eq:UC_lim} is implicit in $r_{n+1}$, requiring a nontrivial numerical solution\footnote{Here the Newton-Raphson method was used, implemented by the \texttt{ newton} function of the \texttt{ numpy} library in Python.}.

Equations \eqref{eq:UC_eq} and \eqref{eq:UC_lim} are made dimensionless by measuring the distances in units of $b$. Considering this, the parameter values are given, following Portela~\cite{portela2007}, as $b = 1$, $R_0 = 30/11$, $a = 8/11$, $l = 8/11$, with $C \approx 0.2167\varepsilon$, associated with the TBR-1 tokamak (where the dimensional scale is $b = 0.11$ meters). Since the aspect ratio $R_0/a$ is large, the poloidal curvature has a relatively small effect, thus the mapping can be well approximated by rewriting it to the Cartesian coordinate system 
\begin{align}
    x &=  \theta, \nonumber \\
    y &=  1 - r/b,
    \label{eq:UC_xy}
\end{align}
where $x$ is defined on the interval $[0, 2\pi]$, while $y = 1$ and $y = 0$ represents the centre of the plasma, and the vessel wall, respectively.

\begin{figure}[!htb]
    \centering
    \includegraphics[width=\linewidth]{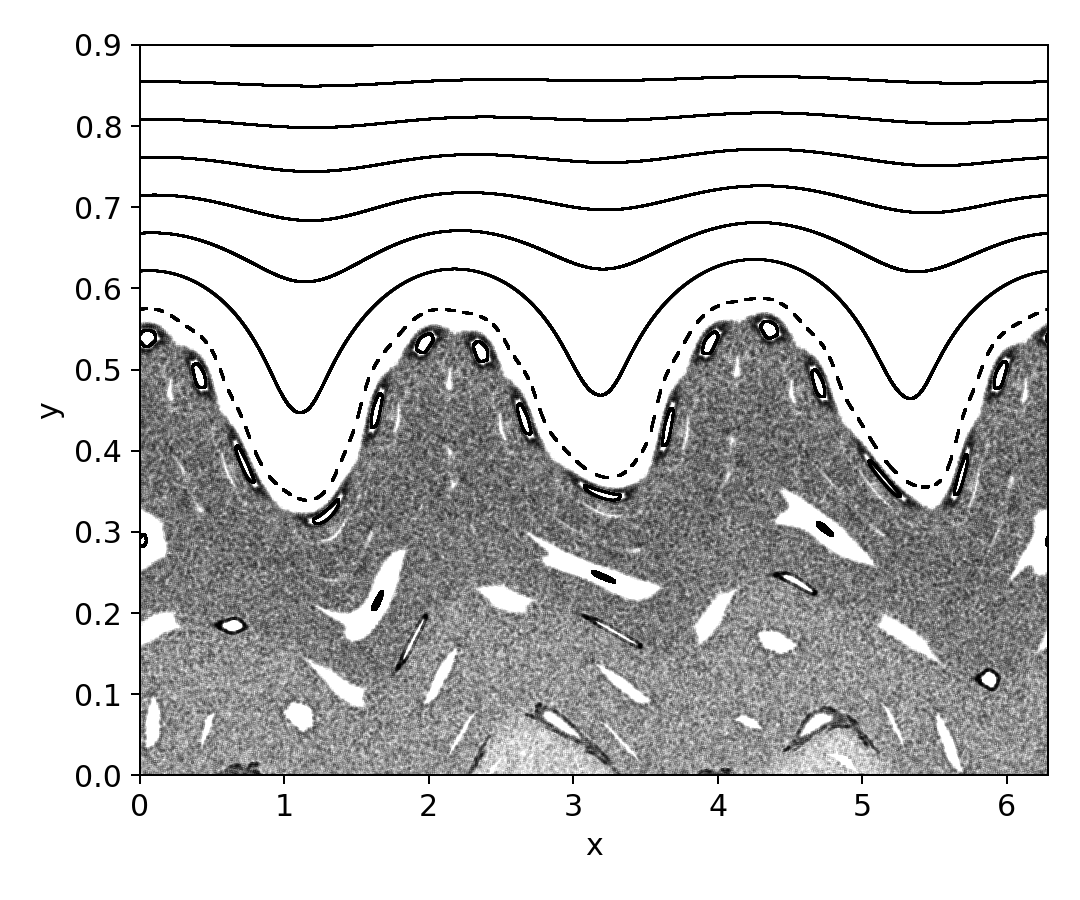}
    \caption{Phase space of the Ullmann-Caldas map in the Cartesian coordinates of \eqref{eq:UC_xy} for $\varepsilon = 0.39$, with particles completing $n = 70\;000$ iterations.}
    \label{fig:poincare_UC}
\end{figure}

Figure \ref{fig:poincare_UC} shows map \eqref{eq:UC_eq}, \eqref{eq:UC_lim} in the Cartesian coordinates \eqref{eq:UC_xy}, with current ratio $\varepsilon = 0.39$, and particles simulated until $n = 70\;000$. The mixed nature of this phase space can be well observed: above around $y = 0.6$ only smooth curves are visible, representing quasi-periodic dynamics in the confined region, while under $y = 0.5$ the chaotic sea dominates, only interrupted by some island chains, appearing as white patches (as no particles were initiated inside them).

\subsection{Transport in the Ullmann-Caldas map}

As our first result, we demonstrate how our proposed model \eqref{eq:transport_function} describes transport in this simple mapping. 
We initiated $N_\text{tot}(0) = 300 \; 000$ initial conditions on 10 horizontal lines of 30 000 particles each, in the phase space of Fig.\ref{fig:poincare_UC}, distributed uniformly in the interval $y = [0.2, 0.4]$. The distribution of particles on the lines was uniform as well. This setup meant that some particles ended up inside islands which then never crossed the escape boundary, which was set to be at a constant $y = 0.1$. The number of particles inside the islands turned out to be $N_\text{island} = 46\;328$, calculated numerically as the number of residual particles at the end of the simulation, when zero particles escape for a considerably long time. 

For easier comparison of the fitting parameters between this case and the JOREK case shown later in section~\ref{sec:JET}, we can hypothetically transform the mapping into the time domain, by assuming that runaway electrons near the speed of light are travelling along the field lines, omitting drift effects. This way one step in the mapping corresponds to the time it takes for a runaway electron to travel around the major radius ($0.3$~meters), which takes roughly $6.28$~nanoseconds.

\begin{figure}[!htb]
    \centering
    \includegraphics[width=\linewidth]{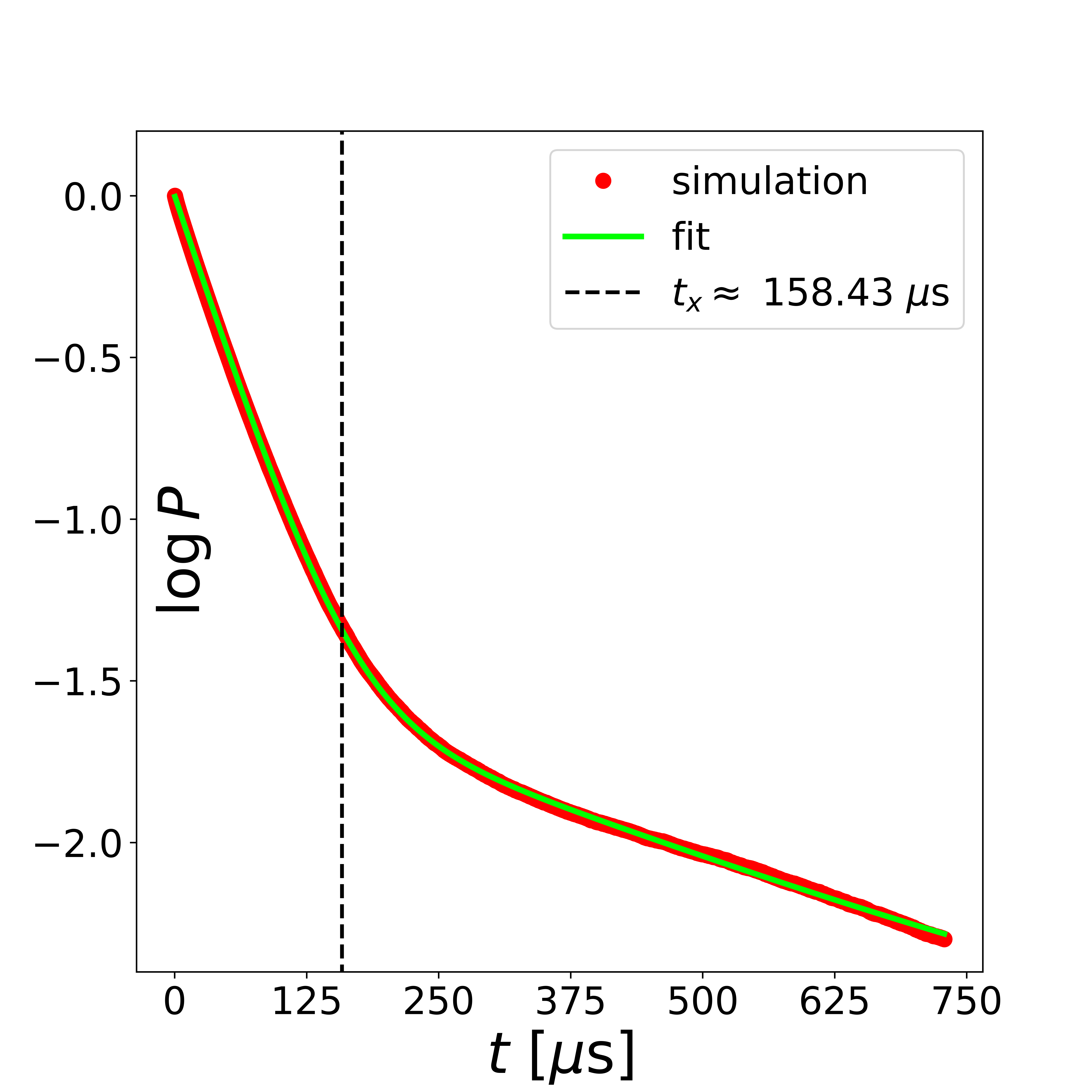}
    \caption{Survival probability of particles for the Ullmann-Caldas model as a function of time (red line) with the transport model \eqref{eq:transport_function} fitted to it (light green line). The fit shows great agreement with the numerical results. The vertical black dashed line indicates the $t_x$ crossover time from \eqref{eq:crossover} .}
    \label{fig:escape_UC}
\end{figure}

Figure \ref{fig:escape_UC} shows the resulting transport curve, i.e. the survival probability $P$ as a function of time $t$. The fitted transport function \eqref{eq:transport_function} (light green line) shows great agreement with the numerical results (red line). The semi-logarithmic scale reveals that the beginning of the process is dominated by the exponential loss of particles, however, from around $t = 250~\mu \text{s}$ the power-law behaviour takes over, and remains to have dominant effect until about $t = 725~\mu \text{s}$, when only a small number -- about 0.49~\% -- of particles (that are not in an island) remain, leading to a breakdown of statistics (not shown).

The fitting when determining the parameter values is detailed in the Supplementary Material. The fitted parameters are
\begin{align}
    &A = (0.87 \pm 5.7\cdot10^{-4}), \nonumber \\
    &\kappa = (0.022 \pm 4.56\cdot10^{-6}) ~\frac{1}{\mu \text{s}}, \nonumber \\
    &\tau = (22.38 \pm 0.11) ~\mu \text{s}, \nonumber \\
    &\alpha = (0.77 \pm 0.0037).
\end{align}
Parameter $A$ is a ratio between the two terms in \eqref{eq:transport_function}, thus its value has to be between 0 and 1 by definition, which is satisfied. The reciprocal of the exponential escape rate $\kappa$ defines the time-scale of the purely diffusive (ergodic) decay as $1/\kappa \approx 45.45~\mu \text{s}$. In earlier models, this time-scale was also associated with the power-law behaviour \cite{altmann2008,altmann2013}. Here, however, just as it was pointed out earlier~\cite{janosi2024}, $1/\kappa$ is notably different from $\tau$. This gives justification for the definition of the distinct parameter $\tau$ associated with the time-scale of the power-law behaviour, a novelty of this transport model. In earlier work~\cite{janosi2024}, parameter $\alpha$ was argued having to be roughly around 1, which is the case here as well.

Using these parameters, the crossover time \eqref{eq:crossover} is determined to be $t_x \approx 158.43~\mu \text{s}$, indicated as a vertical line in Fig.\ref{fig:escape_UC}, where the ratio of particles still inside is $P(t_x) = 4.6\%$.

\section{JET disruption scenario \label{sec:JET}}

The second application of the model presented in this work concerns the chaotic transport of highly energetic electrons in the Joint European Torus (JET)~\cite{mailloux2022overview,Maggi_2024}.
We consider the case of runaway electrons (REs), a subset of the electrons in the plasma which can be accelerated to relativistic energies~\cite{breizman2019physics} during ramp-up or disruptions.
Initially, the REs may still be tied to the magnetic field lines confining the plasma in the device, which form a set of nested closed magnetic surfaces.
Such a post-disruption configuration is usually unstable, however, and perturbations in the magnetic field can lead to changes of the magnetic field topology, in particular related to the formation of magnetic island chains at a given rational surface~\cite{zohm2014magnetohydrodynamic}.
These islands may grow until they saturate, and if at some point two island chains start to overlap, the field between the two surfaces opens up and becomes chaotic within a certain part of the volume, or even the complete plasma domain, in extreme cases.
Understanding this loss of confinement and the details of the ensuing transport is crucial to asses the dynamics of and the potential damage caused by REs, as they are ultimately lost from the confined region and strike the first wall of the device.

The JET discharge in question, JPN \#95135, was targeted at studying REs and the type of magnetohydrodynamic (MHD) instability described earlier.
Some time after the initial disruption and the formation of REs, a large-scale MHD event was observed, which resulted in the prompt flush-out of particles and termination of the RE beam. For details regarding the experiment, refer to the paper by Reux~\cite{reux2021termination}.
Subsequent work using the non-linear extended 3D MHD code JOREK aimed at modelling the instability, including the non-linear phase and the loss of REs caused by transport in the chaotic field topology~\cite{Hoelzl2024recent,Bandaru2021JET}.
While the simulations rely on a fluid model for the REs, meaning transport coefficients need to be prescribed in order to mimic the advection along the field lines, they were able to capture the growth of the magnetic islands and breaking up of the flux surfaces, leading to a near complete loss of the RE population.

A relativistic particle tracing tool has been implemented in JOREK~\cite{Sommariva2018testparticle}, allowing for a more accurate assessment of the RE transport, resolving the full orbit of the particles or using a gyrokinetic approximation.
\Cref{subsec:setup} details how such particle tracing is used to post-process the earlier simulations of Bandaru~\cite{Bandaru2021JET}, deriving the particle loss rate in a given magnetic topology.
The results are then analysed in \Cref{subsec:transport_JET} and compared to the transport model presented in \autoref{eq:transport_function}.

\subsection{Simulation setup}
\label{subsec:setup}

For the full details of the MHD simulation, the reader is referred to the paper by Bandaru~\cite{Bandaru2021JET}.
In this study the fields are taken to be constant in time, corresponding to a time point roughly $81\, \rm \mu s$ into the 3D simulation, when the topology exhibits large chaotic regions as well as clear island-like structures constituting sticky regions.
$2 \cdot 10^{5}$ gyrokinetic RE tracers are initialized at a given normalized poloidal magnetic flux surface $\psi=(\Psi-\Psi_{\rm axis})/(\Psi_{\rm bnd} - \Psi_{\rm axis})$, with $\Psi_{\rm axis}$ and $\Psi_{\rm bnd}$ denoting the poloidal flux at the magnetic axis and the edge of the plasma, respectively.
The particles are then traced until they are either lost from the simulation domain or have completed 20\,000 revolutions around the torus.

\begin{figure}[!htb]
    \centering
    \includegraphics[width=\linewidth]{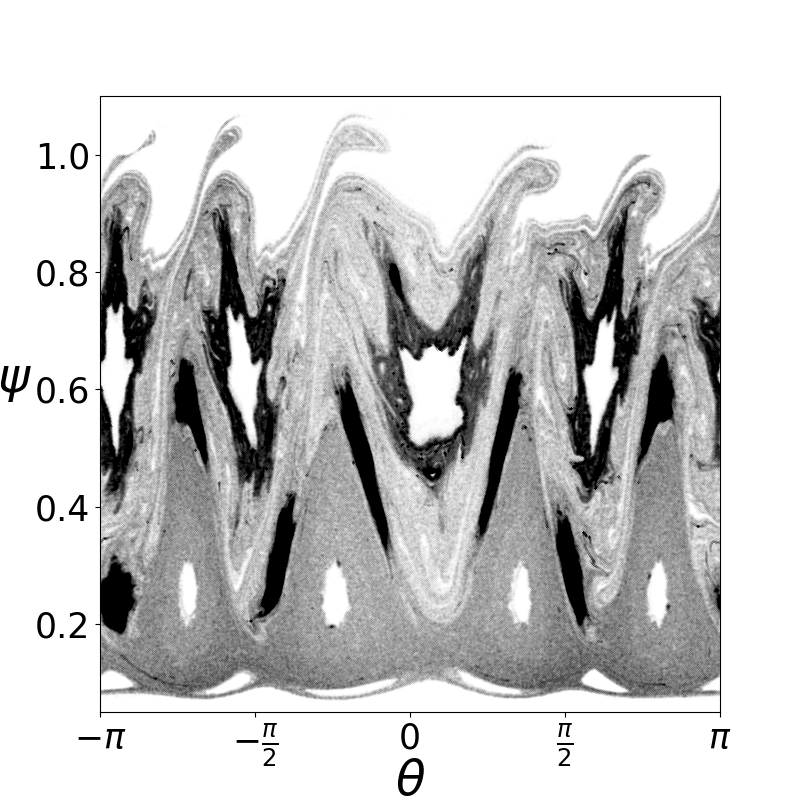}
    \caption{Poloidal surface of the MHD simulation, 
    in terms of the axisymmetric component of the magnetic flux $\psi$ and the poloidal angle $\theta$, with $2\cdot10^5$ particles initiated on the surface $\psi = 0.5$. In this configuration, the confined region is restricted to about $\psi < 0.1$, the nested surfaces are not shown in this figure. The rest of the configuration is dominated by a chaotic sea including two island chains, around which prominent sticky regions can be observed as "denser" regions within the chaotic sea.}
    \label{fig:Poincare_JET}
\end{figure}

An energy of $500~\text{keV}$ is assumed for all REs, with the pitch of the particles set in such a way that $\xi=p_{\parallel}/p=-0.99$, where $p_\parallel$ is the momentum parallel to the magnetic field lines and $p$ the total momentum of the particle.
At this energy, the particles will have a velocity of roughly 86\% of the speed of light ($v=0.86c$), while the curvature drift of the particles remains negligible, implying that the particle orbits closely follow the magnetic field lines.
Each time a particle crosses a given poloidal plane, its location is recorded using as coordinates the axisymmetric ($n=0$) component of $\psi$ and the poloidal angle along the flux surface $\theta$.
The distance travelled for each particle is also recorded, from which the loss time can be derived assuming a constant velocity.
\autoref{fig:Poincare_JET} shows an example of the map obtained when initializing the particles at the $\psi=0.5$ 
surface.
In addition to being prominently chaotic, apparent sticky regions are observed around the two dominant island chains.


\subsection{Transport in the JET scenario}
\label{subsec:transport_JET}

In order to apply the proposed transport function \eqref{eq:transport_function} in the JET disruption scenario, a total of $N_\text{tot} = 1 \; 800 \; 000$ particles were initiated within the Poincaré section displayed in Fig.~\ref{fig:Poincare_JET}, on 9 horizontal lines containing 200 000 particles each in the interval $\Psi = [0.1,0.9]$, distributed in a uniform fashion similar to the case of the Ullmann-Caldas map. The number of particles initiated inside islands this way is $N_\text{island} = 130\;254$. 

\begin{figure}[!htb]
    \centering
    \includegraphics[width=\linewidth]{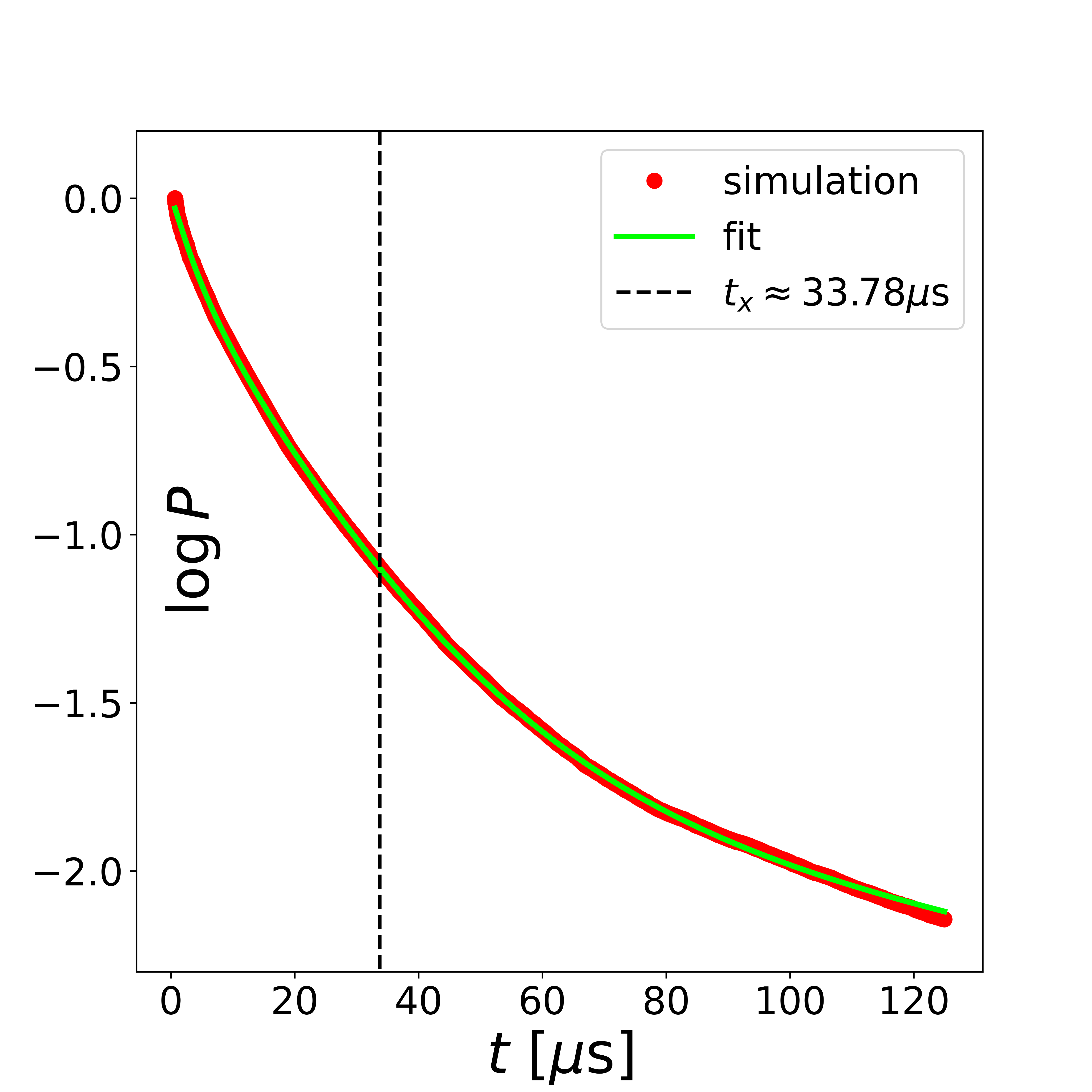}
    \caption{Transport curve in the JET disruption scenario (red) on a semi-logarithmic scale, with the \eqref{eq:transport_function} statistical transport function fitted to it (green), with great agreement. The crossover time $t_x$ is shown as a black vertical dashed line.}
    \label{fig:escape_JET}
\end{figure}

Figure \ref{fig:escape_JET} displays the transport curve (red) with function \eqref{eq:transport_function} fitted to it (green). The fit shows remarkably good agreement with the simulation results. The only slight deviation is at the very end of the shown curve, after which the breakdown in statistics would start, as mentioned in the previous section. This is determined to be around $t = 125\,\rm \mu s$, where only $P(125\,\rm\mu s) \approx 0.71\%$ of particles are not yet lost. 

The fitting parameters (for details, see the Supplementary Material) are obtained as
\begin{align}
    &A = \left(0.39 \pm 4.3\cdot10^{-5}\right), \nonumber \\
    &\kappa = \left(0.068 \pm 1.4\cdot10^{-5} \right) ~ \frac{1}{\mu \text{s}}, \nonumber \\
    &\tau = \left(4.24 \pm 7.1\cdot10^{-4} \right) ~ \mu \text{s}, \nonumber \\
    &\alpha = \left(1.27 \pm 1.7\cdot10^{-4} \right).
\end{align}
The parameters $A$ and $\alpha$ satisfy the conditions set earlier~\cite{janosi2024}. The time-scale of the exponential escape is $1/\kappa \approx 14.7\,\rm \mu s$, which is again different from $\tau$. 
m

The crossover time $t_x \approx 33.78\,\rm\mu s$, determined from \eqref{eq:crossover}, is indicated with a vertical black dashed line, and is somewhat earlier than the obvious change in the character of the transport curve around $t = 70\,\rm \mu s$. The survival probability at the crossover happens to be $P(t_x) \approx 7.9\%$, that is, the escape for the last eight percent of the particles happens predominantly according to a power-law.

The good agreement between the simulation and the theoretical curve in both the Ullmann-Caldas map and the JET disruption scenario shows that the statistical transport function is quite robust in describing the transport of REs in the field line tracing approximation.

\section{Summary}

In this paper, we proposed a model to describe the transport of runaway electrons in perturbed fields, and showed that it describes particle transport both in a mapping model as well as in a 3D nonlinear MHD simulation. The transport model, first introduced by Jánosi and Károlyi~\cite{janosi2024}, is easy to state mathematically, and goes beyond the classical Rechester-Rosenbluth diffusion~\cite{rechester78electron} by incorporating the effect of sticky regions. Studied extensively in chaos theory, these sticky regions slow down transport to a power-law decay of the survival probability instead of an exponential one characteristic of the chaotic regions. The most important assumption of the model is that both chaotic and sticky transport are present at all times, only their relative importance changes.

First, we studied transport in the Ullmann-Caldas map. This is a field line map defined on Poincaré sections, which was originally developed to describe the effect of an ergodic magnetic limiter. The map possesses a safety factor profile similar to those observed in experiments, with geometrical parameters set to those of the TBR-1 tokamak. The transport model fits to the numerical simulations very well, the fitting parameters all being within the expected ranges.

The main result of the paper is the application of the transport model to a disruption scenario of the JET tokamak, simulated with the JOREK code. To do this, a magnetic configuration was chosen in which both chaotic and sticky regions are heavily represented. JOREK's particle tracing tool was used to follow REs, whose energy was set such that the field line tracing approximation was valid. The survival probability of these REs was measured and compared to the theoretical curve, with excellent agreement. By the crossover time, when the effects of the exponential and power-law regimes are equal, around $8\%$ of REs remained and then escaped slowly.

Typical low-fidelity and medium-fidelity disruption models consider the exponential loss model (the diffusive description), but often neglect the fraction of runaway electrons that can remain in the ergodic (chaotic) magnetic field geometry for much longer~\cite{papp2015energetic}. This small fraction of runaway electrons, shown here to be associated with sticky regions, is important to study because it can serve as a seed population for avalanche generation, resulting in a substantial runaway electron current even in the absence of remanent islands in the ergodic field~\cite{Rosenbluth_1997, Boozer_2015, Boozer_2017, breizman2019physics}.


\section*{Acknowledgments}

The authors are grateful to V.~Igochine for fruitful discussions.
This research was supported by the National Research, Development, and Innovation Office (NKFIH) of Hungary, through Grants No. ADVANCED 152888 (D.J., A.H., G.I.P., G.K.), ADVANCED 153324 (D.J., A.H), KDP-2023 C2262591 (D.J., G.V., G.K.) and EK\"OP-25-2-I-ELTE-305 (A.H.). 
This work has been carried out within the framework of the EUROfusion Consortium, funded by the European Union via the Euratom Research and Training Programme (Grant Agreement No 101052200 — EUROfusion). Views and opinions expressed are however those of the author(s) only and do not necessarily reflect those of the European Union, or the European Commission. Neither the European Union nor the European Commission can be held responsible for them.

\bibliographystyle{iopart-num}
\bibliography{refs}

\end{document}